\documentclass[book,numbers]{elsbook-P04}

\usepackage{amsmath,amssymb,amsfonts,amsthm,makeidx,graphicx,colortbl,algorithm,algorithmic}
\usepackage{helvet}
\usepackage{subfigure}
\usepackage{hyperref}

\makeindex

\begin{document}

\Frontmatter

\Mainmatter

\begin{frontmatter}

\chapter[Steganographer Identification]{Steganographer Identification}\label{chap1}

\author*[1]{Hanzhou Wu\footnote{This invited chapter draft is intended as a tutorial introducing the steganographer identification problem in the context of media steganography so that a reader could easily master the basics and key methodologies of SIP. Our focuses are the general technical frameworks, ensemble methods, and feature reduction techniques. No matter what the final decision of this chapter draft will be, if you find any descriptive or technical errors, please do not hesitate to contact me. Errors will be corrected in the revised version. Thank you so much.
}}%

\address[1]{Shanghai University, Shanghai 200444, China}
\address*[1]{\email{h.wu.phd@ieee.org}; \url{https://sites.google.com/view/hzwu}}

\titlemark{Steganographer Identification}
\chaptermark{Steganographer Identification}

\MaxmMiniTocnum{A.1}{3.3.3}{}

\minitoc

\makechaptertitle

\begin{abstract}[Abstract]
Conventional steganalysis detects the presence of steganography within single objects. In the real-world, we may face a complex scenario that one or some of multiple users called \emph{actors} are guilty of using steganography, which is typically defined as the \emph{Steganographer Identification Problem (SIP)}. One might use the conventional steganalysis algorithms to separate stego objects from cover objects and then identify the guilty actors. However, the guilty actors may be lost due to a number of false alarms. To deal with the SIP, most of the state-of-the-arts use unsupervised learning based approaches. In their solutions, each actor holds multiple digital objects, from which a set of feature vectors can be extracted. The well-defined distances between these feature sets are determined to measure the similarity between the corresponding actors. By applying clustering or outlier detection, the most suspicious actor(s) will be judged as the steganographer(s). Though the SIP needs further study, the existing works have good ability to identify the steganographer(s) when non-adaptive steganographic embedding was applied. In this chapter, we will present foundational concepts and review advanced methodologies in SIP. This chapter is self-contained and intended as a tutorial introducing the SIP in the context of media steganography.
\end{abstract}

\begin{keywords}[Keywords:]
Steganographer identification, batch steganography, pooled steganalysis, clustering, anomaly detection, unsupervised learning, social networks.
\end{keywords}

\end{frontmatter}%

\section{Introduction}
Steganography \cite{fridrich:book} hides from untrusted parties the fact that any secret is being communicated. It can be realized by embedding the secret message into an innocent object (e.g., digital image) called \emph{cover}. The resultant object called \emph{stego} will be sent to the trusted receiver who can perfectly retrieve the secret message from the stego according to a key. A most important requirement for any secure steganographic system is that it should be impossible for an eavesdropper to distinguish between ordinary objects and objects that contain secret information. From the perspective of the opponent, steganalysis reveals the presence of hidden messages. The objective of steganography and steganalysis can be described as follows. As shown in Figure 1.1, Alice wants to send message \textbf{m} into a randomly sampled genuine cover \textbf{x}. She utilizes a key to drive the embedding procedure to generate the stego \textbf{y}, which is sent through the \emph{insecure} channel and inspected by Eve's detection algorithm, the steganalyzer. Eve has to gather evidence about the type of communication from a single output of the steganalyzer and decide whether or not to cut the communication channel. If Bob receives \textbf{y}, he extracts \textbf{m} with the key.

\begin{figure}
  \centering
  \includegraphics[width=4.5in]{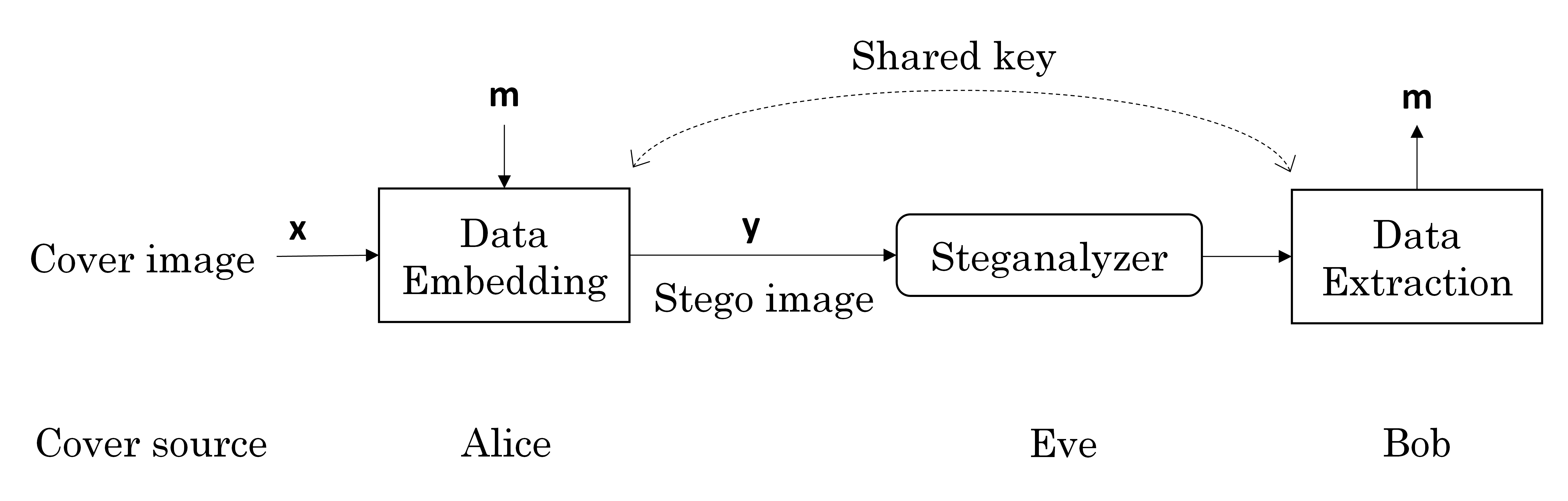}\\
  \caption{Model of steganographic communication channel.}\label{fig1-1}
\end{figure}

There are two common principles for designing steganographic systems \cite{fridrich:gibbs}. One is model-preserving steganography, aiming to preserve the selected model of the cover source during data embedding, e.g., the histogram. The other one treats steganography as a \emph{rate-distortion} optimization task. For a payload, we hope to minimize the data embedding distortion. In other words, we want to embed as many message bits as possible for a distortion. Current advanced steganographic algorithms first assign costs of changing individual elements based on their local neighborhood and then hide the payload using coding techniques with (near) minimal total embedding cost \cite{fridrich:stcs}.

Steganalysis has been greatly developed as well. As shown in Figure 1.2, the contemporary approach to steganalysis involves three components. It extracts steganalytic features from objects; it supplies training sets of cover and stego objects; and it runs a machine learning algorithm (e.g., SVM) on the training data. This creates a decision function for novel objects, classifying them as cover or stego. Since steganography generally modifies noise-like components of the cover, advanced feature extractors aim to mine the statistical difference in high-frequency areas for detection.

\begin{figure}
  \centering
  \includegraphics[width=4.5in]{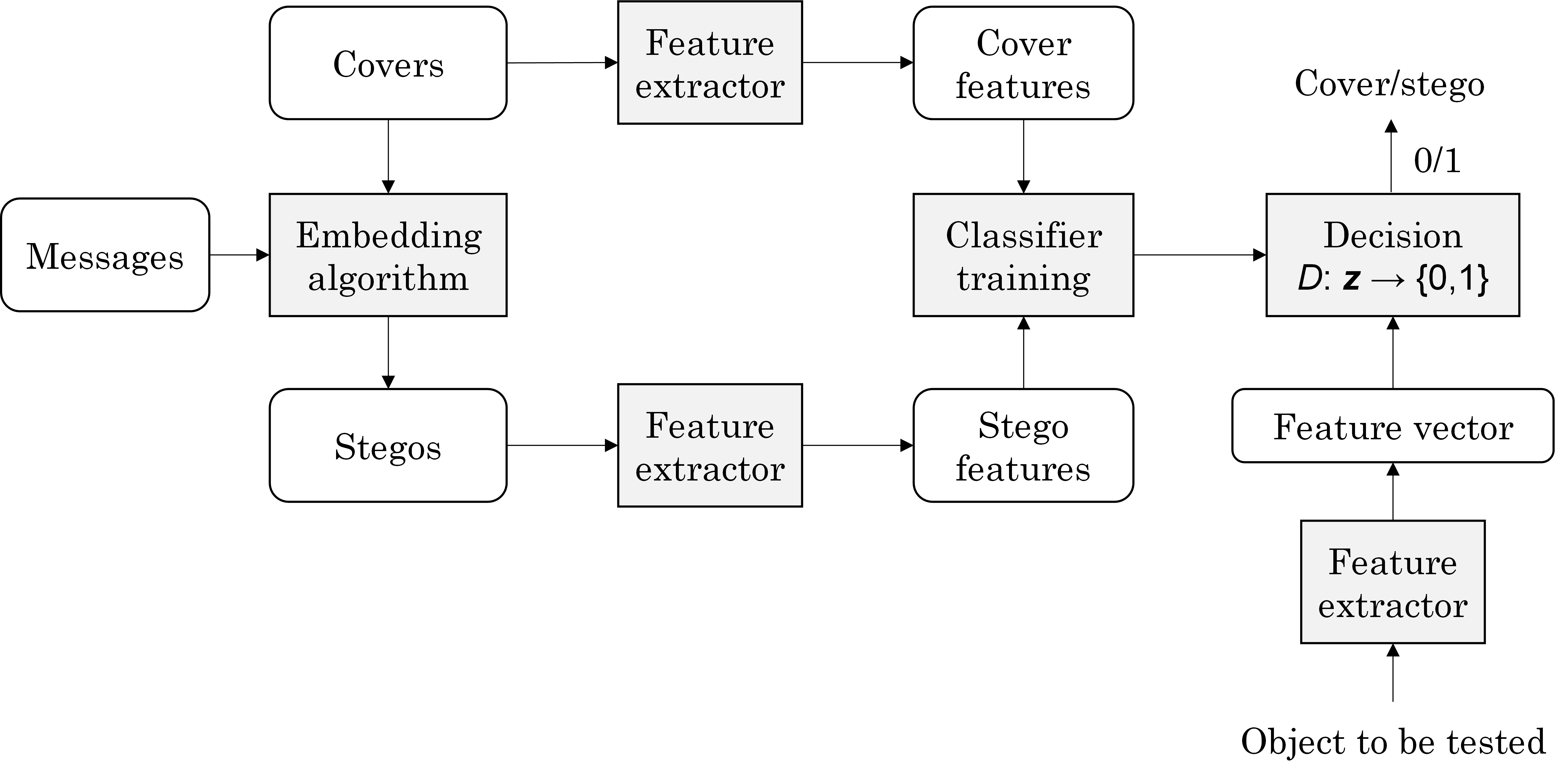}\\
  \caption{Steganalysis as a binary classification problem.}\label{fig1-2}
\end{figure}

In traditional steganalysis, each cover object is treated in isolation, which is implicit in the vast majority of the literature, e.g., they separate cover objects from stego objects, working on each object individually. However, such a restriction is unrealistic: In practice, we may face a complex scenario that multiple network actors send a set of media files while one or some of them are using steganography. To make matters even more complicated, an actor who is guilty of performing steganography will probably behave innocently some of the time, mixing the stego objects with genuine covers \cite{batch2006}. Therefore, any steganalyst will surely have to consider multiple actors, and each will transmit multiple objects, which is referred to as the \emph{Steganographer Identification Problem (SIP)}. One might use traditional steganalysis to find stegos and then identify the guilty actors. However, the guilty actors may be lost due to a number of false alarms. Because, most traditional steganalysis algorithms involve a binary classification algorithm, which has to be trained on large sets of covers and stegos. In practice, the cover source of the actors will not be identical to that used for training by the steganalyst, so as well as being computationally demanding the training itself becomes a possible source of error if an innocent actor is falsely accused because of training model mismatch \cite{Ker2011, Ker2012}.

It therefore requires us to design reliable detectors for the aforementioned complex scenario. Fortunately, though the SIP needs further investment, there are some effective works reported in the literature. Figure 1.3 shows an example to identify the guilty actor from multiple actors by unsupervised learning. Each actor holds multiple images, from which multiple feature sets can be extracted. Each feature set can be regarded as a ``point''. By using unsupervised learning based method such as clustering, the most suspicious point, i.e., feature set D, can be found out. Accordingly, Kim will be judged as the guilty actor. Following this line, in this chapter, we will introduce the foundational concepts and detail the methodologies of SIP. We hope this chapter could help readers to master both basic concepts and advanced technologies for SIP.

The structure of this chapter is organized as follows. In Section 1.2, we present the primary concepts and techniques. Then, in Section 1.3, we show two mainstream technical frameworks for SIP. In Section 1.4, we introduce the ensemble and dimensionality reduction techniques for improving identification performance. Finally, we conclude this chapter in Section 1.5.

\begin{figure}
  \centering
  \includegraphics[width=4.5in]{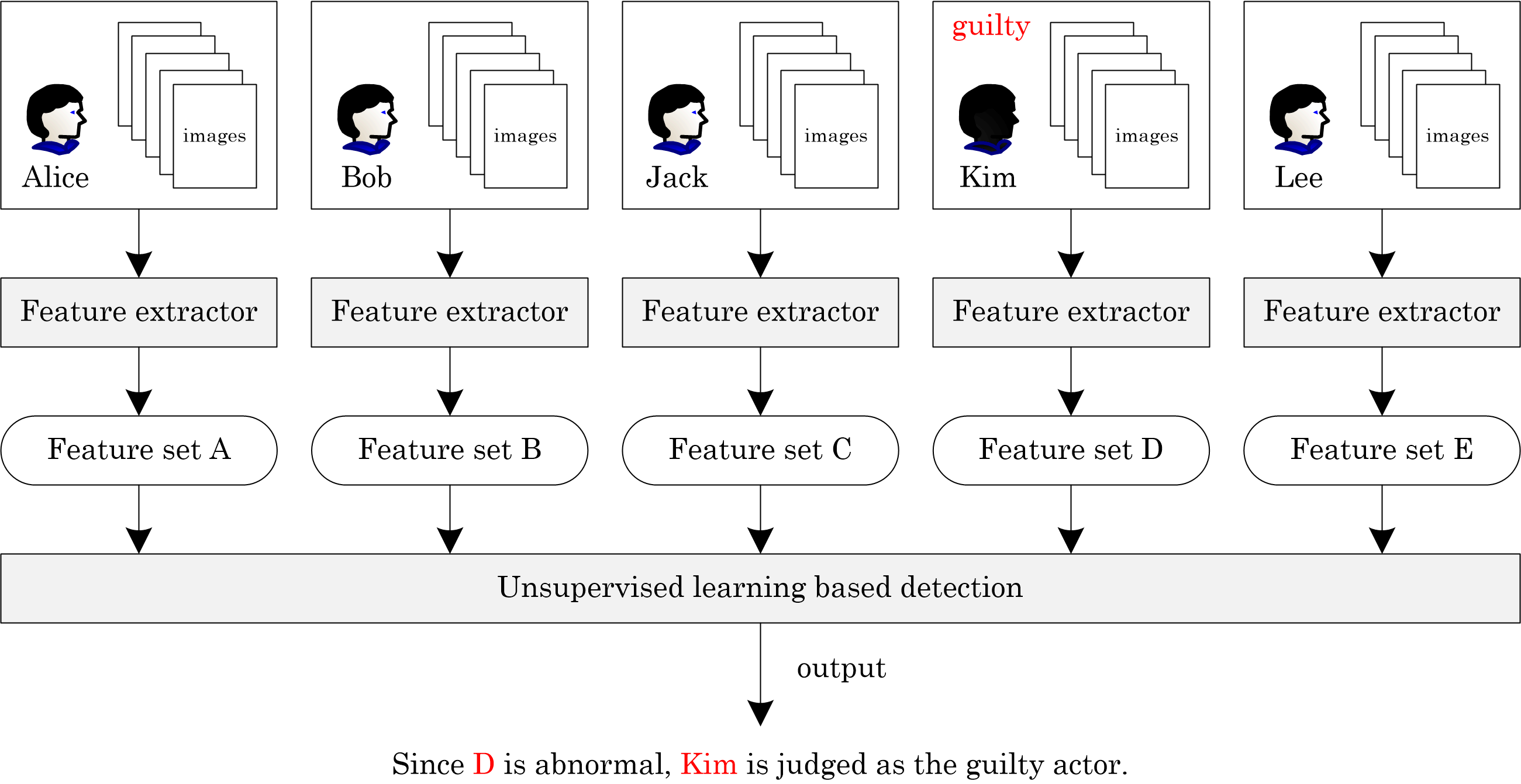}\\
  \caption{An example to identify the guilty actor by unsupervised learning.}\label{fig1-3}
\end{figure}

\section{Primary Concepts and Techniques}
\subsection{JPEG Compression}
JPEG is a common method of lossy compression for digital images. The degree of compression can be adjusted, allowing a selectable tradeoff between storage size and image quality. Unless mentioned, we will use JPEG images as the covers for steganography throughout this chapter because JPEG image is one of the most popular type of cover over social networks. We here briefly review the JPEG compression procedure in JPEG images.

The JPEG compression algorithm is designed specifically for the human eye. It exploits the biological properties of human sight: we are more sensitive to the illuminocity of color, rather than the chromatric value of an image, and we are not particularly sensitive to high-frequency content in images. The encoding process of the JPEG compression algorithm has five main basic steps:

\begin{enumerate}
  \item The representation of the colors in the image is converted from RGB to $\text{Y}\text{C}_\text{B}\text{C}_\text{R}$, having one luminance $\text{Y}$, and two chrominances $\text{C}_\text{B}$ and $\text{C}_\text{R}$.
  \item The resolution of the chroma data is reduced. This reflects the fact that eyes are less sensitive to fine color details than to fine brightness details.
  \item The image is split into blocks of $8\times 8$ pixels, and for each block, each of the Y, $\text{C}_\text{B}$, and $\text{C}_\text{R}$ data undergoes the discrete cosine transform (DCT).
  \item The amplitudes of the frequency components are quantized. The quality setting of the encoder affects to what extent the resolution of each frequency component is reduced. The magnitudes of the high-frequency components are stored with a lower accuracy than the low-frequency components.
  \item The resulting data for all $8\times 8$ blocks is further compressed with a lossless algorithm, a variant of Huffman encoding.
\end{enumerate}

The decoding process reverses these steps, except the quantization because it is irreversible. Y, $\text{C}_\text{B}$, and $\text{C}_\text{R}$ can be determined as:
\begin{align}
\left\{\begin{matrix}
\begin{split}
\text{Y}=&~0.299\text{R}+0.587\text{G}+0.114\text{B},\\
\text{C}_\text{B}=&~0.564(\text{B}-\text{Y}),\\
\text{C}_\text{R}=&~0.713(\text{R}-\text{Y}).
\end{split}
\end{matrix}\right.
\end{align}

For a pixel-block sized $8\times 8$, the DCT can be described as:
\begin{align}
F(u,v)=\frac{1}{4}C(u)C(v)\sum_{x=0}^{7}\sum_{y=0}^{7}f(x,y)\text{cos}\frac{(2x+1)u\pi}{16}\text{cos}\frac{(2y+1)v\pi}{16},
\end{align}
\begin{align}
f(x,y)=\frac{1}{4}\sum_{u=0}^{7}\sum_{v=0}^{7}C(u)C(v)F(u,v)\text{cos}\frac{(2x+1)u\pi}{16}\text{cos}\frac{(2y+1)v\pi}{16},
\end{align}
where $C(z)=1/\sqrt{2}$ for $z=0$; otherwise, $C(z)=1$. Here, $f(u,v)$ denotes the value at position $(u,v)$ to be transformed, and, $F(u,v)$ is the DCT coefficient.

Once DCT is finished, JPEG moves on to quantization, in which the less important DCT coefficients are wiped out. For quantization, we need a quantization table $T\in \mathbb{Z}^{8\times 8}$. After quantization, $F(u,v)$ will be updated as $\lfloor \frac{F(u,v)}{T(u,v)} \rfloor$.

\subsection{JPEG Steganalysis Features}
Each actor's JPEG images are represented by low dimensional feature vectors. We call them as JPEG steganalysis features. In this section, we review two JPEG feature extractors since they will be used latter. However, it is always free for us to design other efficient feature extractors for JPEG steganalysis.

\subsubsection{PEV-274 Features}
T. Pevn$\acute{\text{y}}$ and J. Fridrich \cite{PEV274} merged the DCT features and Markov features to produce a 274-D feature vector called PEV-274 for JPEG steganalysis.

Let the luminance of a JPEG file be represented with a DCT coefficient array $d_{i,j}(k)$, $i,j=1,...,8$, $k=1,...,n_B$, where $d_{i,j}(k)$ denotes the $(i,j)$-th quantized DCT coefficient in the $k$-th block (there are $n_B$ blocks). Let \textbf{H} be the histogram of all $64\times n_B$ luminance DCT coefficients, i.e., $\textbf{H} = (H_L, ..., H_R)$, where $L=\text{min}_{i,j,k}d_{i,j}(k)$ and $R=\text{max}_{i,j,k}d_{i,j}(k)$. Let $\textbf{h}^{i,j} = (h_L^{i,j}, ..., h_R^{i,j})$ be a histogram of coefficients for DCT mode $(i,j)$. Let the dual histograms with $8\times 8$ matrices be $\textbf{g}_{i,j}^d = \sum_{k=1}^{n_B}\delta (d, d_{i,j}(k))$, where $i,j=1,...,8$, $d=-5,...,5$, and $\delta (x, y) = 1$ if $x = y$ and 0 otherwise. The inter-block dependency among DCT coefficients can be captured. First, we find the variation $V = $:
\begin{align*}
\begin{split}
&\frac{\sum_{i,j=1}^{8}\sum_{k=1}^{|\textbf{I}_r|-1}|d_{i,j}(\textbf{I}_r(k))-d_{i,j}(\textbf{I}_r(k+1)) |}{|\textbf{I}_r|+|\textbf{I}_c|}+\\
&~~~~~~~~~~~~~~~~~~~~~~~~~~~~~~\frac{\sum_{i,j=1}^{8}\sum_{k=1}^{|\textbf{I}_c|-1}|d_{i,j}(\textbf{I}_c(k))-d_{i,j}(\textbf{I}_c(k+1))|}{|\textbf{I}_r|+|\textbf{I}_c|},
\end{split}
\end{align*}
where $\textbf{I}_r$ and $\textbf{I}_c$ denote the vectors of block indices $1,...,n_B$ while scanning the image by rows and by columns, respectively. Then, two scalars calculated from the decompressed JPEG image representing an integral measure of inter-block dependency over all DCT modes over the whole image are collected:
\begin{align*}
B_\alpha = \frac{\sum_{i=1}^{\lfloor (M-1)/8 \rfloor}\sum_{j=1}^{N}|\textbf{c}_{8i,j}-\textbf{c}_{8i+1,j}|^\alpha + \sum_{j=1}^{\lfloor (N-1)/8 \rfloor}\sum_{i=1}^{M}|\textbf{c}_{i,8j}-\textbf{c}_{i,8j+1}|^\alpha}{N\lfloor (M-1)/8 \rfloor + M \lfloor (N-1)/8 \rfloor },
\end{align*}
where $M$ and $N$ are height and width, $\textbf{c}_{i,j}$ is the grayscale value of the decompressed JPEG image, $\alpha = 1, 2$. We can also collect calibrated co-occurrence features, e.g., $\textbf{C}_{0,0}(J_1) - \textbf{C}_{0,0}(J_2)$, where $\textbf{C}_{s,t} = \sum_{i,j=1}^{8}E_{i,j}$ and $E_{i,j}$ equals:
\begin{align*}
\begin{split}
&\frac{\sum_{k=1}^{|\textbf{I}_r-1|}\delta(s,d_{i,j}(\textbf{I}_r(k)))\delta(t,d_{i,j}(\textbf{I}_r(k+1)))}{|\textbf{I}_r|+|\textbf{I}_c|}+\\
&~~~~~~~~~~~~~~~~~~~~~~~~~~~~~~\frac{\sum_{k=1}^{|\textbf{I}_c|-1}\delta(s,d_{i,j}(\textbf{I}_c(k)))\delta(t,d_{i,j}(\textbf{I}_c(k+1)))}{|\textbf{I}_r|+|\textbf{I}_c|}.
\end{split}
\end{align*}

$J_1$ is the stego image, and $J_2$ is its calibrated version. Calibration estimates macroscopic properties of the cover from the stego. During calibration, $J_1$ is decompressed to the spatial domain, cropped by a few pixels in two directions, and compressed again with the same quantization matrix as $J_1$. The newly obtained $J_2$ has the most macroscopic features similar to the original cover. This is because the cropped image is visually similar to the original image. Moreover, the cropping operation brings the $8\times 8$ DCT grid ``out of sync'' with the previous compression, which effectively suppresses the influence of the previous JPEG compression and the embedding changes.

The above analysis allows us to construct 193-D DCT features. In detail, we first collect 66-D histogram features, i.e.,
\begin{align}
\textbf{H}_l(J_1) - \textbf{H}_l(J_2),l\in \{-5, -4, ..., 5\},
\end{align}
and for all $(i,j)\in \mathcal{L}$ = $\{(1,2)$, $(2,1)$, $(3,1)$, $(2,2)$, $(1,3)\}$,
\begin{align}
\textbf{h}_{l}^{i,j}(J_1) - \textbf{h}_{l}^{i,j}(J_2), l\in \{-5, -4, ..., 5\}.
\end{align}

For dual histograms $\textbf{g}^d, d\in \{-5, -4, ..., 5\}$, we determine the difference of the 9 lowest AC modes, i.e.,
\begin{align}
\textbf{g}_{i,j}^d(J_1) - \textbf{g}_{i,j}^d(J_2),
\end{align}
where $(i,j)\in \{(2,1),(3,1),(4,1),(1,2),(2,2),(3,2),(1,3),(2,3),(1,4)\}$. It allows us to collect 99-D DCT features. For the co-occurrence matrix, the central elements in the range $[-2,2]\times [-2,2]$ are used, yielding to 25-D features
\begin{align}
\textbf{C}_{s,t}(J_1) - \textbf{C}_{s,t}(J_2), (s,t)\in [-2,2]\times [-2,2].
\end{align}

By further collecting $V$, $B_1$, $B_2$, we have 193-D DCT features. The Markov feature set proposed in \cite{ShiJPEGMarkov2006} models the differences between absolute values of neighboring DCT coefficients as a Markov process. The calculation starts by forming the matrix $F(u,v)$ of absolute values of DCT coefficients in the image. The DCT coefficients in $F(u,v)$ are arranged in the same way as pixels in the image by replacing each $8\times 8$ block of pixels with the corresponding block of DCT coefficients Four difference arrays can be calculated along four directions: horizontal, vertical, diagonal, and minor diagonal, i.e.,
\begin{align}
\left\{\begin{matrix}
\begin{split}
F_h(u,v) &= F(u,v) - F(u+1,v),\\
F_v(u,v) &= F(u,v) - F(u,v+1),\\
F_d(u,v) &= F(u,v) - F(u+1,v+1),\\
F_m(u,v) &= F(u+1,v) - F(u,v+1).
\end{split}
\end{matrix}\right.
\end{align}

From these difference arrays, four transition probability matrices $\textbf{M}_h$, $\textbf{M}_v$, $\textbf{M}_d$, $\textbf{M}_m$ are constructed as:
\begin{align}
\left\{\begin{matrix}
\begin{split}
\textbf{M}_h(i,j) &= \frac{\sum_{u,v}\delta(F_h(u,v)=i,F_h(u+1,v)=j)}{\sum_{u,v}\delta(F_h(u,v)=i)},\\
\textbf{M}_v(i,j) &= \frac{\sum_{u,v}\delta(F_v(u,v)=i,F_v(u,v+1)=j)}{\sum_{u,v}\delta(F_v(u,v)=i)},\\
\textbf{M}_d(i,j) &= \frac{\sum_{u,v}\delta(F_d(u,v)=i,F_d(u+1,v+1)=j)}{\sum_{u,v}\delta(F_d(u,v)=i)},\\
\textbf{M}_m(i,j) &= \frac{\sum_{u,v}\delta(F_m(u+1,v)=i,F_m(u,v+1)=j)}{\sum_{u,v}\delta(F_m(u,v)=i)}.
\end{split}
\end{matrix}\right.
\end{align}

If the matrices are taken directly as features, the dimensionality would be too large. By using the central $[-4, 4]$ portion of the matrices, the dimension is then $4\times 9^2 = 324$. Obviously, the above Markov features can be calibrated, i.e., $\textbf{M}^{(c)} = \textbf{M}(J_1) - \textbf{M}(J_2)$. The dimension of the calibrated Markov feature set, $\{\textbf{M}_h^{(c)}, \textbf{M}_v^{(c)}, \textbf{M}_d^{(c)}, \textbf{M}_m^{(c)}\}$, remains the same as its original version. PEV-274 uses the average $\bar{\textbf{M}} = (\textbf{M}_h^{(c)} + \textbf{M}_v^{(c)} + \textbf{M}_d^{(c)} + \textbf{M}_m^{(c)}) / 4$, which has a dimension of 81. Accordingly, for PEV-274, the dimension is $193+81=274$.

\subsubsection{LI-250 Features}
Li \emph{et al}. \cite{Li2016} proposed high-order joint features for SIP. The motivation is that, most JPEG steganographic methods would change the correlation of neighboring coefficients due to the modification of DCT coefficiants. Thus, Markov transition probability will be affected with the data embedding operation. However, Markov transition probabilities may be incapable of fully exploiting the changes in DCT coefficients and may not completely capture the neighboring joint relation. Thus, Li \emph{et al}. developed a 250-D features, by employing joint density matrices, for pooled JPEG steganalysis. The high-order joint features are driven from high-order joint density matrices of DCT coefficients, which indicate the intra-block and inter-block dependencies of JPEG images. The features include two parts: one part is generated from mean joint density matrices of intra-block, while the other is derived from the mean joint density matrices of inter-block. The details of extracting the features are as follows.

Denoting the quantized DCT coefficients in the blocks of JPEG image as $c_{m,n}(u,v)$, where $m\in [1, M]$ and $n \in [1, N]$ are the indices of blocks, and $u$ and $v$ indicate the positions of coefficients in each block. The intra-block joint density matrices along horizontal, vertical, and diagonal directions, denoted by $F_{ia}^{(h)}(x,y,z)$, $F_{ia}^{(v)}(x,y,z)$, $F_{ia}^{(d)}(x,y,z)$, are calculated by:
\begin{align*}
\frac{\sum_{m,n,u,v}\delta(|c_{m,n}(u,v)|=x, |c_{m,n}(u,v+1)|=y, |c_{m,n}(u,v+2)|=z)}{48MN},
\end{align*}
\begin{align*}
\frac{\sum_{m,n,u,v}\delta(|c_{m,n}(u,v)|=x, |c_{m,n}(u+1,v)|=y, |c_{m,n}(u+2,v)|=z)}{48MN},
\end{align*}
\begin{align*}
\frac{\sum_{m,n,u,v}\delta(|c_{m,n}(u,v)|=x, |c_{m,n}(u+1,v+1)|=y, |c_{m,n}(u+2,v+2)|=z)}{36MN}.
\end{align*}

The intra-block joint features $F_{ia}(x,y,z)$ are then calculated by:
\begin{align}
F_{ia}(x,y,z) = \frac{1}{3}\{F_{ia}^{(h)}(x,y,z) + F_{ia}^{(v)}(x,y,z) + F_{ia}^{(d)}(x,y,z)\},
\end{align}
where $x,y,z\in [0, 4]$. Thus, the dimension of intra-block joint features is 125. Similarly, the inter-block second-order joint density matrices along the three directions, denoted by $F_{ir}^{(h)}(x,y,z)$, $F_{ir}^{(v)}(x,y,z)$, $F_{ir}^{(d)}(x,y,z)$, can be calculated respectively by:
\begin{align*}
\frac{\sum_{m,n,u,v}\delta(|c_{m,n}(u,v)|=x, |c_{m,n+1}(u,v)|=y, |c_{m,n+2}(u,v)|=z)}{64M(N-2)},
\end{align*}
\begin{align*}
\frac{\sum_{m,n,u,v}\delta(|c_{m,n}(u,v)|=x, |c_{m+1,n}(u,v)|=y, |c_{m+2,n}(u,v)|=z)}{64(M-2)N},
\end{align*}
\begin{align*}
\frac{\sum_{m,n,u,v}\delta(|c_{m,n}(u,v)|=x, |c_{m+1,n+1}(u,v)|=y, |c_{m+2,n+2}(u,v)|=z)}{64(M-2)(N-2)}.
\end{align*}

The inter-block joint features $F_{ir}(x,y,z)$ are calculated by
\begin{align}
F_{ir}(x,y,z) = \frac{1}{3}\{F_{ir}^{(h)}(x,y,z) + F_{ir}^{(v)}(x,y,z) + F_{ir}^{(d)}(x,y,z)\},
\end{align}
where $x,y,z\in [0,4]$. Thus, the dimension of inter-block joint features is 125. By combining the intra-block features and the inter-block features, we can get 250-D high-order joint features, which will be used for SIP later.

\subsection{Batch Steganography and Pooled Steganalysis}
In SIP, each actor holds multiple digital objects. Normal actors take no action to their own digital objects. While, for a guilty actor, i.e., the steganographer, he distributes a secret payload into the objects held by himself. It involves the concept of \emph{batch steganography} \cite{batch2006}, which can be described as follows: given a total of $n$ cover objects, the steganographer hides data in $m\leq n$ of them and leaves the other covers alone. Obviously, to secure steganography, how to best spread payload between multiple covers is a critical problem to the steganographer. For a steganalysis expert, how to pool evidence from multiple objects of suspicion is a key topic, which is called \emph{pooled steganalysis} \cite{batch2006}. The methods for SIP can be regarded as pooled steganalysis.

Recalling the goal of batch steganography: the steganographer wants to spread a message of length $L$ among images $(I_1,I_2,...,I_n)$ with secure capacities $(c_1,c_2,...,c_n)$ using a steganographic embedding algorithm for individual images. He needs to determine the message fragment lengths $(l_1,l_2,...,l_n)$, with $L=\sum_{i=1}^{n}l_i$, to be embedded into the images. There are four common strategies reported in the literature \cite{KerBatchRealWorld}: max-greedy, max-random, linear and even. In the max-greedy strategy, the steganographer embeds the message into the fewest possible number of covers. Assuming that the images are ordered by capacity $c_1> c_2> ... > c_n$, it leads to the following message lengths: $l_i=c_i, \forall i\in [1,m-1]$, $l_m=L-\sum_{i=1}^{m-1}l_i$, and $l_i=0, \forall i\in [m+1, n]$. The max-random strategy is the same as max-greedy, except that the images to be embedded are chosen in a random order. In the linear strategy, the message is distributed into all images proportionately to their capacity, i.e., $l_i=\frac{c_iL}{\sum_{j=1}^{n}c_j}$. For the even strategy, the secret message is distributed evenly, i.e., $l_i=L/n$. The existing works \cite{batch2006, KerGame, KerCapacitySPL} point that, for the steganographer, the optimal behaviour is likely to be extreme concentration of payload into as few covers as possible, or the opposite in which payload is spread as thinly as possible. However, these theoretical results could not be confirmed without practical pooled steganalyzers to test against.

The aforementioned strategies require the steganographer to estimate the capacities. Ker \emph{et al}. \cite{KerBatchRealWorld} presented a method to estimating the maximum message length for each available image for practical steganography. They first query the implementation of an algorithm to provide an initial estimate of the maximum message length. This is done either by embedding a very short message into the given message, or asking for information about a given image. Once having an initial estimate, they try to embed a randomly string of this length. If the embedding fails, the estimate of the capacity is decreased by ten bytes and the procedure is repeated. Otherwise, they deem the current estimate of the capacity as the final. It is pointed that, the capacity for an image is actually not easy to be well-defined in terms of security. One may exploit steganalysis for estimating the secure capacity, rather than simply estimating the maximum embeddable payload as the capacity.

\subsection{Agglomerative Clustering}
The hierarchical cluster analysis \cite{ClusterMethods} seeks to build a hierarchy of clusters, allowing us to partition a set of objects such that ``similar'' objects can be clustered. As a type of hierarchical cluster analysis, agglomerative clustering is such a ``bottom-up'' approach that each object starts in its own cluster, and pairs of clusters are merged as one moves up the hierarchy. As shown in \textbf{Algorithm 1}, the basic agglomerative clustering maintains an ``active set'' of clusters (initially, all objects are belonging to the active set) and at each stage the nearest two clusters are merged. When two clusters are merged, they are removed from the active set and their union is added to the active set. It iterates until there is only one cluster in the active set. The tree is formed by keeping track of which clusters were merged. All that is needed is a method to compute a distance between two clusters, for which there are a number of options. We introduce several distance measures below.

\begin{algorithm}[!t]
 \caption{Basic Agglomerative Clustering}
 \begin{algorithmic}[1]
	\STATE Collect objects $\{x_i\}_{i=1}^n$, and set distance measure $D(\cdot,\cdot)$.
    \STATE $\mathcal{A} \leftarrow \emptyset$ \hfill $\triangleright$ Active set starts out empty.
    \FOR {each $i \in [1,n]$}
        \STATE $\mathcal{A}\leftarrow \mathcal{A} \cup \{\{x_i\}\}$\hfill $\triangleright$ Add each object as its own cluster.\
    \ENDFOR
    \STATE $\mathcal{T} \leftarrow \mathcal{A}$ \hfill $\triangleright$ Store the tree as a sequence of merges.
	\WHILE {$|\mathcal{A}|>1$}
	   \STATE $\mathcal{G}_1^*,\mathcal{G}_2^*\leftarrow \underset{\mathcal{G}_1,\mathcal{G}_2\in \mathcal{A}}{\text{arg min}}~D(\mathcal{G}_1,\mathcal{G}_2)$ \hfill $\triangleright$ Choose pair in $\mathcal{A}$ with best distance.
        \STATE $\mathcal{A}\leftarrow\mathcal{A}\setminus\{\mathcal{G}_1^*\}\setminus\{\mathcal{G}_2^*\}$ \hfill $\triangleright$ Remove both from active set.
        \STATE $\mathcal{A}\leftarrow\mathcal{A}\cup\{\mathcal{G}_1^*\cup\mathcal{G}_2^*\}$ \hfill $\triangleright$ Add union to active set.
        \STATE $\mathcal{T}\leftarrow\mathcal{T}\cup\{\mathcal{G}_1^*\cup\mathcal{G}_2^*\}$ \hfill $\triangleright$ Add union to tree.
	\ENDWHILE
	\RETURN Tree $\mathcal{T}$
 \end{algorithmic}
\end{algorithm}

Let us write $d(x, y)$ for the distance between two objects $x$ and $y$, and $D(X, Y)$ for the distance between two clusters $X$ and $Y$. The single linkage uses the distance between the nearest points in the two clusters:
\begin{align}
D_\text{SL} = \underset{x\in X, y\in Y}{\text{min}}~d(x,y).
\end{align}

The complete linkage uses the furthest points:
\begin{align}
D_\text{CL} = \underset{x\in X, y\in Y}{\text{max}}~d(x,y).
\end{align}

The single linkage can cause long chains of clusters, whereas complete linkage prefers compact clusters; other agglomerative clustering algorithms are intermediate, including centroid clustering
\begin{align}
D_\text{CEN} = \frac{1}{|X|\cdot |Y|}\sum_{x\in X}\sum_{y\in Y}d(x,y),
\end{align}
and average linkage
\begin{align}
D_\text{AL} = \frac{1}{|X\cup Y|^2-|X\cup Y|}\underset{u,v\in X\cup Y,u\neq v}{\sum\sum}~d(u,v).
\end{align}

The input to the basic agglomerative clustering is a distance matrix between objects. The output can be displayed in a dendrogram, a tree of the successive cluster agglomerations using ``height'' to indicate the distance between clusters being merged.

\subsection{Local Outlier Factor}
The local outlier factor (LOF) \cite{LOFpaper} is an outlier detection method by measuring the local deviation of a sample point with respect to its neighbours, which requires a distance measure $d: \mathcal{X}\times \mathcal{X}\mapsto \mathbb{R}$. Assuming that, we have a set of feature points $C$ (vectors). We expect to estimate the degree of outlying of the point $p\in C$. The LOF uses a parameter $k\in \mathbb{N}$, with $1<k<|C|$, specifying the number of nearest neighbors.

In LOF, the $k$-distance of $p$, denoted by $d_k(p)$, is defined as a distance between $p$ and $o\in C\setminus \{p\}$, such that:
\begin{itemize}
  \item for at least $k$ points $o'\in C\setminus\{p\}$, it holds $d(p,o')\leq d_k(p)$, and
  \item for at most $k-1$ points $o'\in C\setminus\{p\}$, it holds $d(p,o')< d_k(p)$.
\end{itemize}

The $k$-neighborhood of $p$ is defined as $N_k(p)=\{o\in C| d(p,o)\leq d_k(p)\}$. Note that, the number of elements in the $k$-th neighborhood, i.e., $|N_k(p)|$, can be greater than $k$. The reachability distance of $p$ w.r.t. $o$ is defined as $r_k(p,o)=\text{max}\{d_k(o),d(p,o)\}$. The local reachability density of $p$ is defined as an inverse of the average reachability distance of $p$ to all points $o$ in its $k$-neighborhood, i.e.,
\begin{align}
\text{LRD}_k(p)=\frac{|N_k(p)|}{\sum_{o\in N_k(p)}r_k(p,o)}.
\end{align}

The LOF method uses reachability distance to estimate density of the probability at point $p$. It also estimates the average density of the probability of all points in the $k$-neighborhood, $o\in N_k(p)$, and compares these quantities. The level of outlying is therefore adaptive with respect to the closest neighborhood. Thus, the LOF value of $p$ is finally defined as:
\begin{align}
\text{LOF}_k(p)=\frac{1}{|N_k(p)|}\sum_{o\in N_k(p)}\frac{\text{LRD}_k(o)}{\text{LRD}_k(p)}.
\end{align}

The $\text{LOF}_k(p)$ captures degree to which $p$ can be called an outlier. A high value indicates that $p$ is an outlier, since its reachability density, $\text{LRD}_k(p)$, is smaller in comparison to the local reachability density of its neighbors.

\subsection{Maximum Mean Discrepancy}
In SIP, we need a measure of distance between two actors. Since each actor has a set of feature vectors, we have to design a distance measure for two sets of feature vectors. The maximum mean discrepancy (MMD) \cite{MMDthesis, MMDpaper, MMDsurvey} has been empirically shown to be quite effective for distance measurement. It can be used for SIP. Given observations $X=\{\textbf{x}_i\}_{i=1}^{|X|}$ and $Y=\{\textbf{y}_i\}_{i=1}^{|Y|}$, which are i.i.d. drawn from $p(\textbf{x})$ and $q(\textbf{y})$ defined on $\mathbb{R}^d$, let $\mathcal{F}$ be a class of functions $f:\mathbb{R}^d\mapsto \mathbb{R}$, the MMD and its empirical estimate are:
\begin{equation}
\text{MMD}[\mathcal{F},p,q] = \underset{f\in \mathcal{F}}{\text{sup}}~\mathbb{E}_{\textbf{x}\sim p(\textbf{x})}f(\textbf{x})-\mathbb{E}_{\textbf{y}\sim q(\textbf{y})}f(\textbf{y}),
\end{equation}
\begin{equation}
\text{MMD}[\mathcal{F},X,Y] = \underset{f\in \mathcal{F}}{\text{sup}}~\frac{1}{|X|}\sum_{\textbf{x}\in X}f(\textbf{x})-\frac{1}{|Y|}\sum_{\textbf{y}\in Y}f(\textbf{y}).
\end{equation}

Usually, $\mathcal{F}$ is selected as a unit ball in a universal RKHS $\mathcal{H}$ defined on compact metric space $\mathbb{R}^d$ with kernel $k(\cdot,\cdot)$ and feature mapping $\phi(\cdot)$. The Gaussian and Laplacian kernels are universal. It has been proven that,
\begin{equation}
\text{MMD}^2[\mathcal{F},p,q] = \left \|\mathbb{E}_{\textbf{x}\sim p(\textbf{x})}\phi(\textbf{x})-\mathbb{E}_{\textbf{y}\sim q(\textbf{y})}\phi(\textbf{y}) \right \|_{\mathcal{H}}^2.
\end{equation}
An \emph{unbiased} estimate of MMD is:
\begin{equation}
\text{MMD}[\mathcal{F},X,Y] = \left (\frac{1}{|X|^2-|X|}\sum_{i\neq j}h[i,j]\right )^{1/2},
\end{equation}
where $|X| = |Y|$ is assumed and
\begin{equation}
h[i,j] = k(\textbf{x}_i,\textbf{x}_j) + k(\textbf{y}_i,\textbf{y}_j) - k(\textbf{x}_i,\textbf{y}_j) - k(\textbf{x}_j,\textbf{y}_i).
\end{equation}

For any two sets, we can use the unbiased estimate of MMD to measure their distance. However, it is noted that, when a set has only one feature vector, we cannot use MMD since its value always equals zero. In this case, one may use Euclidean metric, i.e., $d(\textbf{x},\textbf{y})=||\textbf{x}-\textbf{y}||_{2}$, or other metrics. A kernel function is required when to use MMD. There are choices for $k$. For example, the linear kernel which is simply a scalar product
\begin{equation}
k(\textbf{x},\textbf{y}) = \textbf{x}\cdot\textbf{y},
\end{equation}
and the Gaussian kernel is defined as
\begin{equation}
k(\textbf{x},\textbf{y}) = \text{exp}(-\gamma ||\textbf{x}-\textbf{y}||^2),
\end{equation}
where $\gamma$ is a parameter. Typically, $\gamma$ is set to $\eta^2$, where $\eta$ is the median of the $L_2$-distance between features in the set of images being considered: this means that the exponents are, on average, close to -1.

\section{General Frameworks}
In this section, we review two general frameworks used in SIP. It can be said that, most of the present state-of-the-arts base on the two frameworks. Both are proposed by Ker \emph{et al.}, which can be found in references \cite{Ker2011, Ker2012, Ker2014}.

\subsection{Clustering-based Detection}
Suppose that, multiple actors each will transmit multiple objects, all of which have been intercepted. Each actor may use multiple sources of objects. For simplicity, we assume that, each actor has just one source of objects, but that these sources could be different from each other. For each actor, we compute the feature vectors from the objects held by this actor. We can think of this as many clouds of points in the feature space, one cloud for each actor. A guilty actor is one who using steganography in (some of) their transmitted objects. We aim to identify guilty actors if their clouds of feature vectors stand out, in some way, from the normal actors'.

Applying steganalysis to each object individually might be valuable, but it is likely that any guilty party could be lost in a crowd of false positives. If we use a trained model for the objects to be tested, it will produce the mismatching problem in practice. It is possible to perform clustering on the individual objects, which may provide little information. It inspires us to consider each actor's objects as a whole to characterize their source. In this way, we are to cluster the actors, hoping to separate an innocent majority from a guilty minority. Here, clustering the actors is equivalent to clustering the clouds mentioned above. Each cloud consists of a set of feature vectors.

Thus, we combine the hierarchical clustering with the MMD distance measure for SIP. The distances between actors are defined as the MMD distance between the sets of feature vectors. If the features have been well-chosen so that the difference between actors' sources is less than the difference between guilty and normal actors, the final agglomeration will be between two clusters, the normal and the guilty. Thus, we can extract a list of suspected guilty actors from the cluster dendrogram. As well as depending on the features, the accuracy of this method will also depend on the proportion of objects which guilty actors embed in, and the amount they embed. Such detector assumes that, the majority of actors is normal, and tries to identify guilty actors as an outlier cluster.

Pre-processing the directly extracted features is necessary to guarantee the accuracy of the detection method since the raw features have different scales. Feature normalization is a good choice and other preprocessing methods may be suitable as well such as principal component transformation. By normalization, we mean linear scaling of features such that each column of the data matrix $\widehat{\text{X}}$ has zero mean and unit variance. This means that
\begin{equation}
\frac{1}{nl}\sum_{i=1}^{nl}\widehat{\text{X}}_{ir}=0,\forall r,
\end{equation}
and
\begin{equation}
\frac{1}{nl}\sum_{i=1}^{nl}\widehat{\text{X}}_{ir}^2=1,\forall r,
\end{equation}
where $n$ is the number of actors, and $l$ for the number of objects held by each actor. The preprocessing enables the distance measure to be more meaningful and not significantly affected by noisy components.

Mathematically, let $A=\{a_1,a_2,...,a_n\}$ and $S(a_i)=\{\textbf{I}_1^{(i)},\textbf{I}_2^{(i)},...,\textbf{I}_m^{(i)}\}$ respectively represent the actors and the objects held by actor $a_i$. Based on the aforementioned analysis, we describe the detailed steps for the (hierarchical) clustering based framework as follows.

\begin{enumerate}
  \item For each actor $a_i$, $1\leq i\leq n$, with a well-designed feature extractor $\mathcal{E}_\text{fea}$, we extract the feature vectors from the objects in $S(a_i)$, which is denoted by $F(a_i)=\{\textbf{f}_1^{(i)},\textbf{f}_2^{(i)},...,\textbf{f}_m^{(i)}\}$.
  \item For each actor $a_i$, $1\leq i\leq n$, normalizing $F(a_i)$ by the above method.
  \item For any two different actors $a_i$ and $a_j$, $1\leq i<j\leq n$, determine the MMD distance between normalized $F(a_i)$ and normalized $F(a_j)$.
  \item Apply hierarchical clustering, and collect two clusters $\mathcal{C}_1$ and $\mathcal{C}_2$ at the final stage of merging. The actors belonging to the cluster with a smaller size are considered as the guilty actors.
\end{enumerate}

\begin{figure}
  \centering
  \includegraphics[width=4.5in]{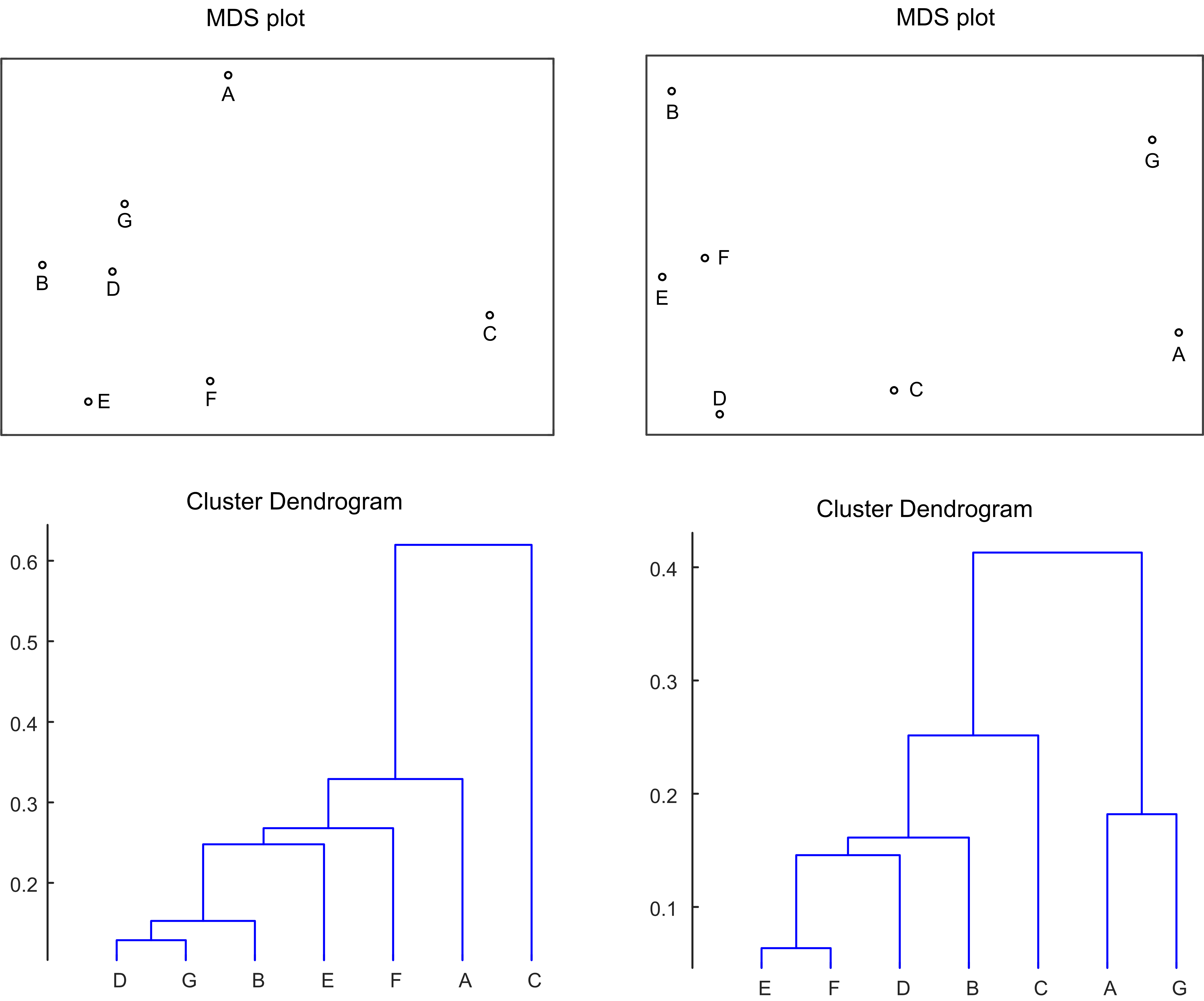}\\
  \caption{Examples for MDS representation and cluster dendrogram.}\label{fig1-4}
\end{figure}

Assuming that, we want to find $k\leq |\mathcal{C}_1|+|\mathcal{C}_2|$ suspicious actors, and $|\mathcal{C}_1|\leq |\mathcal{C}_2|$. If $k\leq |\mathcal{C}_1|$, we randomly choose $k$ actors from $\mathcal{C}_1$ as the guilty actors. Otherwise, all actors in $\mathcal{C}_1$ and $k-|\mathcal{C}_1|$ randomly-selected actors in $\mathcal{C}_2$ are judged as guilty ones. For better understanding, we take Figure 1.4 for explanation. In Figure 1.4, the above means MDS representation of actors, and the below shows cluster dendrograms, from which the final two clusters are encircled in the MDS plot. The left example implies $\mathcal{C}_1 = \{$C$\}$ and $\mathcal{C}_2 = \{$A, B, D, E, F, G$\}$. For the right example, we have $\mathcal{C}_1 = \{$A, G$\}$ and $\mathcal{C}_2 = \{$B, C, D, E, F$\}$. Suppose that, there has only one guilty actor. Then, for the left example, C will be judged as the suspicious actor. For the right example, either A or G will be judged as the steganographer. If there are three guilty actors, for the right example, $\{$A, G$\}$ and an actor randomly chosen from $\mathcal{C}_2$ can be selected out as the guilty actors.

\subsection{Outlier-based Detection}
The advantage of the hierarchical clustering based framework is that, by pooling objects from each actor together, the method improves the signal-to-noise ratio compared with working on individual objects. This leads to better accuracy in the identification of guilty actor. The rationale for using a clustering algorithm was that the group of guilty actors should form a distinct cluster, that would be well separated from the normal actors. In reality, differences in actors' cover sources means that there is already variation in feature vectors between difference actors. The steganographic embedding may exhibit itself by a shift in the feature vectors in some direction(s). Consequently, the guilty actors could be represented by outliers, and thus allows us to switch our attention from clustering to outlier detection.

A lot of outlier detection methods can be found in the literature. The local outlier factor (LOF) method would be a good choice as: (a) it requires only the pairwise distances between points, (b) it is not tied to any particular application domain, (c) it directly provides a measure of how much an outlier each point is, and (d) it relies on single hyper-parameter. We use LOF for detection here. However, we point out that, it is always free for us to apply any effective outlier detection method for SIP.

\begin{algorithm}[!t]
 \caption{LOF based detection method for SIP}
\begin{algorithmic}[1]
\renewcommand{\algorithmicrequire}{\textbf{Input:}}
\renewcommand{\algorithmicensure}{\textbf{Output:}}
\REQUIRE $A=\{a_1,...,a_n\}$, $S(a_i)=\{\textbf{I}_1^{(i)},...,\textbf{I}_m^{(i)}\},i\in [1,n].$
\ENSURE A ranking list $\textbf{r}$. \hfill $\triangleright$ $r_k\in\textbf{r}$ reveals the $k$-th most suspicious actor.
\STATE Extract feature vectors for each actor
\STATE Preprocess feature vectors by normalization  \hfill $\triangleright$ Other methods also work.
\STATE Apply LOF to the $n$ points (i.e., sets of normalized feature vectors)
\STATE Return a ranking list $\textbf{r}$ based on the LOF values \hfill $\triangleright$ $\textbf{r}=(r_1,r_2,...,r_n)$.
\end{algorithmic}
\end{algorithm}

We have introduced the concept of LOF in the previous section. Here, we directly present the pseudo-code of using LOF for SIP, which is shown in \textbf{Algorithm 2}. It is seen that, the steps are quite similar to that of the hierarchical clustering based framework. The LOF method does not provide a threshold, above which the element should be considered as an outlier. Such method has an advantage. Namely, the LOF values do not depend on the absolute values of distances, i.e., it is scale invariant. It is worth mentioning that, by switching to outlier detection, we have overcome another shortcoming of the clustering approach, which was the lack of any direct measure of ``being an outlier''. If we know that we want to identify a single guilty, we can take the one with the largest LOF value; if we want to identify a short list of the $k$ most suspicious actors, we can choose those with top-$k$ LOF values.

\subsection{Performance Evaluation and Analysis}
We report experimental results in \cite{Ker2011, Ker2014} for performance analysis.
\subsubsection{Clustering-based Detection}
Suppose that, there are $n$ actors, each of whom has $m$ images, and we want to identify a single guilty actor from them. The actors are simulated by images taken from $n$ different cameras, all are JPEG compressed with an identical quality factor (QF). nsF5 \cite{nsF5} is used as the steganographic algorithm, which is an improved version of F5 \cite{F5}. F5 is a steganographic algorithm that aims to preserve the shape of the histogram of quantized DCT coefficients, making it similar to that of the cover image. The message is embedded by changing the absolute values of DCT coefficients toward zero. F5 is also the first algorithm to use matrix embedding, which is a coding scheme that increases embedding efficiency measured by the number of bits embedded per embedding change. As its variant, nsF5 uses the same type of embedding changes as F5. To avoid introducing more zeros, however, nsF5 uses wet paper codes \cite{wpc, wpc2} with improved efficiency.

Each image is represented by a 274-D feature vector called PEV-274, designed for JPEG steganalysis and previously shown to be effective against nsF5. However, it is admitted that, nsF5 is detectable by modern steganalysis features. The hierarchical clustering technique with MMD distance between actors is used. The expectation is that the final agglomeration should be between one guilty actor and a cluster of $n-1$ normal actors. It is pointed that, there is no training in the paradigm, and it does not assume any knowledge of the actors' cameras or the embedding algorithm.

\begin{table*}
\centering
\caption{Confusion matrix for identifying one guilty actor: 0.25 bpnc, 25\%.}
\begin{tabular}{c c c c c c c c}
\\\hline
& \multicolumn{7}{c}{\textbf{Identified actor}} \\
\cmidrule(l){2-6}
\textbf{Guilty actor} & A & B & C & D & E & F & G\\
\midrule
A & 100 &   0 &   0 &   0 &   0 &   0 &   0\\
B &   0 & 100 &   0 &   0 &   0 &   0 &   0\\
C &   0 &   0 & 100 &   0 &   0 &   0 &   0\\
D &   0 &   1 &   5 &  94 &   0 &   0 &   0\\
E &   0 &   9 &   7 &   0 &  84 &   0 &   0\\
F &   0 &   9 &  14 &   0 &   0 &  77 &   0\\
G &   0 &  16 &   7 &   0 &   0 &   0 &  77\\
\hline
\end{tabular}
\end{table*}

\begin{table*}
\centering
\caption{Confusion matrix for identifying one guilty actor: 0.3 bpnc, 30\%.}
\begin{tabular}{c c c c c c c c}
\\\hline
& \multicolumn{7}{c}{\textbf{Identified actor}} \\
\cmidrule(l){2-6}
\textbf{Guilty actor} & A & B & C & D & E & F & G\\
\midrule
A & 100 &   0 &   0 &   0 &   0 &   0 &   0\\
B &   0 & 100 &   0 &   0 &   0 &   0 &   0\\
C &   0 &   0 & 100 &   0 &   0 &   0 &   0\\
D &   0 &   0 &   0 & 100 &   0 &   0 &   0\\
E &   0 &   0 &   0 &   0 & 100 &   0 &   0\\
F &   0 &   0 &   1 &   0 &   0 &  99 &   0\\
G &   0 &   0 &   0 &   0 &   0 &   0 & 100\\
\hline
\end{tabular}
\end{table*}

There are $n = 7$ actors sending JPEG images. Each actor converts the RAW photos to JPEG QF=80 before transmitting them. Each actor uses images of a constant size (the default resolution of the camera), but the image size does vary from actor to actor. Tables 1.1 and 1.2 show the confusion matrices for the identification of one guilty actor, with each experiment repeated 100 times per guilty actor. Each actor transmits $m=50$ JPEG images: Table 1.1, the guilty actor embeds 0.25 bits per non-zero coefficient (bpnc) in 25\% of his images; Table 1.2, the guilty actor embeds 0.3 bpnc in 30\% of his images. The single linkage and linear MMD are used here. The identification of the guilty party is performed by cutting the dendrogram at the final agglomeration, which has been described previously. It can be observed that, the overall accuracy are 90.3\% and 99.9\%, respectively. Performance falls of sharply for smaller payloads since the guilty actors fade into the normal cluster and they cannot be identified accurately.

\begin{table*}
\centering
\caption{Overall accuracy due to different parameter settings.}
\begin{tabular}{c c c c c c c}
\\\hline
& & & \multicolumn{4}{c}{\textbf{Guilty actor's proportion of}} \\
& & & \multicolumn{4}{c}{\textbf{stego images @ payload per image}} \\
\cmidrule(l){2-6}
\textbf{MMD} & \textbf{Cluster} & $m$ & 10\% @ & 30\% @ & 50\% @ & 70\% @\\
\textbf{kernel} & \textbf{linkage} &  & 0.1 bpnc & 0.3 bpnc & 0.5 bpnc & 0.7 bpnc\\
\midrule
Linear & Single & 200 & 21.3\% & 100.0\% & 100.0\% & 100.0\%\\
Linear & Single & 100 & 18.0\% & 100.0\% & 100.0\% & 100.0\%\\
Linear & Single &  50 & 17.7\% &  99.9\% & 100.0\% & 100.0\%\\
Linear & Single &  20 & 14.1\% &  96.7\% & 100.0\% & 100.0\%\\
Linear & Single &  10 & 15.9\% &  86.4\% &  99.9\% & 100.0\%\\
\hline
Linear & Centroid & 200 & 17.0\% & 100.0\% & 100.0\% & 100.0\%\\
Linear & Centroid & 100 & 16.6\% & 100.0\% & 100.0\% & 100.0\%\\
Linear & Centroid &  50 & 16.3\% & 100.0\% & 100.0\% & 100.0\%\\
Linear & Centroid &  20 & 13.4\% &  97.0\% & 100.0\% & 100.0\%\\
Linear & Centroid &  10 & 15.4\% &  87.1\% &  99.9\% & 100.0\%\\
\hline
Linear & Average & 200 & 17.0\% & 100.0\% & 100.0\% & 100.0\%\\
Linear & Average & 100 & 16.4\% & 100.0\% & 100.0\% & 100.0\%\\
Linear & Average &  50 & 15.7\% & 100.0\% & 100.0\% & 100.0\%\\
Linear & Average &  20 & 14.3\% &  97.0\% & 100.0\% & 100.0\%\\
Linear & Average &  10 & 16.6\% &  87.1\% &  99.9\% & 100.0\%\\
\hline
Linear & Complete & 200 & 15.4\% & 100.0\% & 100.0\% & 100.0\%\\
Linear & Complete & 100 & 14.4\% & 100.0\% & 100.0\% & 100.0\%\\
Linear & Complete &  50 & 13.9\% &  99.3\% & 100.0\% & 100.0\%\\
Linear & Complete &  20 & 15.6\% &  94.9\% & 100.0\% & 100.0\%\\
Linear & Complete &  10 & 16.6\% &  84.0\% &  99.9\% & 100.0\%\\
\hline
Gaussian & Centroid & 200 & 14.0\% &  92.0\% & 100.0\% & 100.0\%\\
Gaussian & Centroid & 100 & 13.7\% &  87.4\% & 100.0\% & 100.0\%\\
Gaussian & Centroid &  50 & 12.6\% &  76.1\% & 100.0\% & 100.0\%\\
Gaussian & Centroid &  20 & 12.9\% &  60.1\% &  99.7\% & 100.0\%\\
Gaussian & Centroid &  10 & 15.3\% &  40.9\% &  98.4\% & 100.0\%\\
\hline
\end{tabular}
\end{table*}

Experiments are conducted with both linear and Gaussian MMD. Table 1.3 shows the experimental results due to different parameter settings. It can be found that, the clustering based on linear MMD worked better than that based on Gaussian MMD. Also, the choice of agglomeration algorithm did not make a substantial difference. As expected, there is a dependency on $m$: more evidence allows the detector to make more accurate detection from smaller payloads. When $m = 200$, the linear MMD combined with any linkage achieved perfectly accurate detection of the guilty parity with payloads as small as 0.3 bpnc embedded in only 30\% of the images.

\subsubsection{Outlier-based Detection}
Compared to clustering-based detection, outlier-based detection allows us to consider up to thousands actors each transmitting up to hundreds images. In \cite{Ker2014}, in each experiment, a randomly selected subset of 4000 actors and 200 images for each actor was used. The images were obtained from a social network site, which automatically resized large images, to approximately 1Mpix, and then JPEG compressed with QF = 85, at the time of crawling. Exactly one guilty actor was simulated by nsF5. Four different embedding strategies, i.e., max-greedy, max-random, linear and even, were used to insert $Lp$ bits to the images held by the guilty actor. $L$ is the total number of non-zero coefficients in the images. $p$ is the number of bits per non-zero coefficient. The PEV-274 features were extracted from the $nm$ JPEG images, and then normalized. The MMD distance between each pair of actors can be then determined. Finally, the LOF detection method was applied to compute the LOF score for each actor, which allowed us to identify the most suspicious actor that will be judged as the guilty actor.

\begin{figure}
  \subfigure[$n = m = 100$]{
    \begin{minipage}[t]{0.45\linewidth}
    \centering
    \includegraphics[width=2.1in]{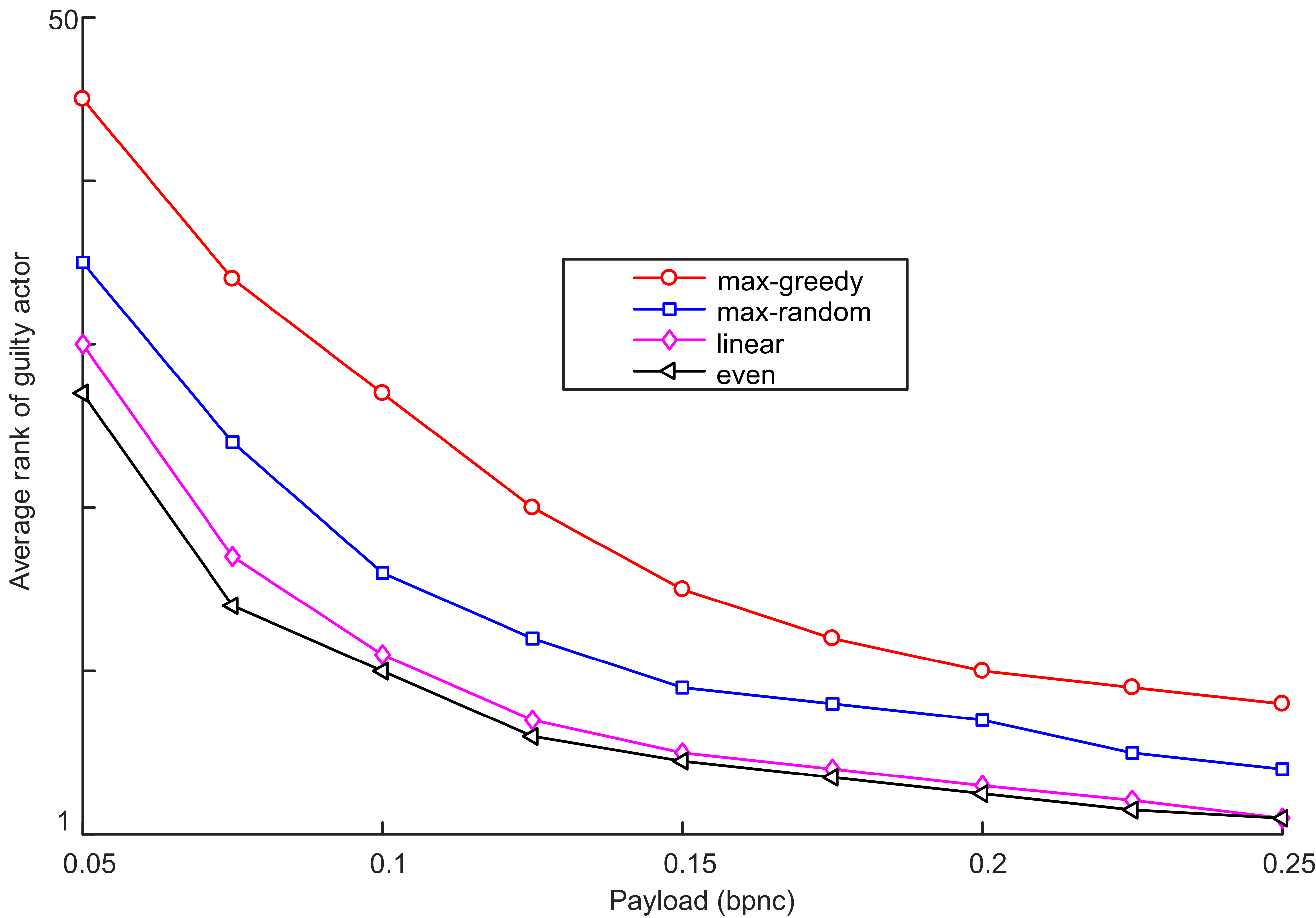}
    \end{minipage}
  }
  \subfigure[$n = 400, m = 100$]{
    \begin{minipage}[t]{0.45\linewidth}
    \centering
    \includegraphics[width=2.1in]{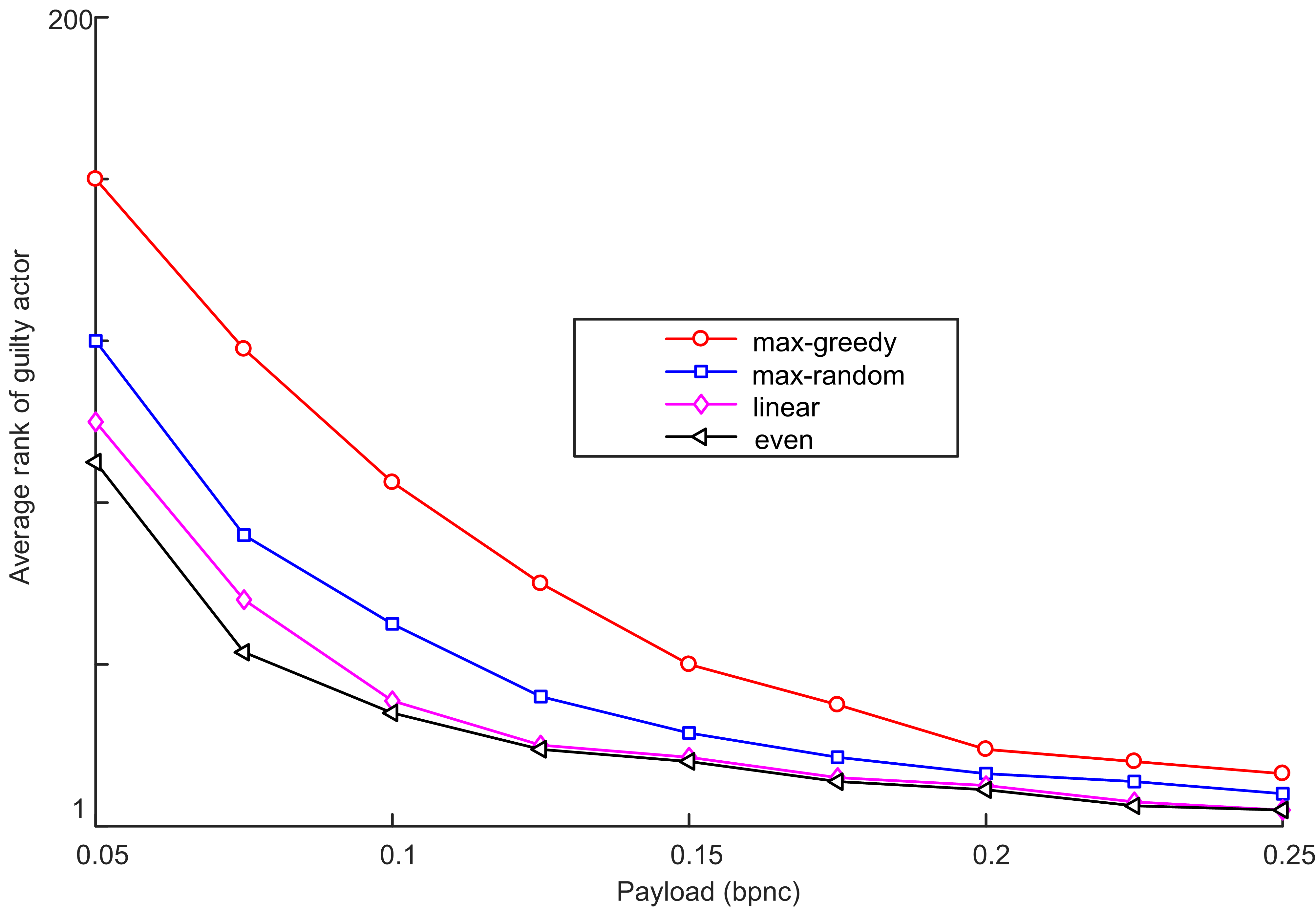}
    \end{minipage}
  }
  \caption{Detection performance due to different parameters.}\label{fig1-5}
\end{figure}

For each combination of parameters, each experiment was repeated 500 times with a different selection of actors and guilty actor. The \emph{average rank of the guilty actor} was used as a benchmark to reflect how well the guilty actor was identified. An average rank of one corresponds to perfect detection, i.e., the guilty actor is always ranked most suspicious. And, an average rank of $\frac{n+1}{2}$ corresponds to random guessing. We do not expect a universal, unsupervised detector to achieve perfect accuracy, but instead hope that, it provides intelligence by ranking a truly guilty actor among the top 10\% of all actors. In addition, for the MMD distance, the centroid `kernel' (a variant of linear kernel, please refer to the Appendix of \cite{Ker2014}) was used. And, for the LOF detection, a parameter $k$ was set to 10. Figure 1.5 show the detection performance due to different parameters. It can be observed that, the max-greedy strategy is the most secure for the embedder: the average rank of the guilty actor is higher in all payloads. The second most secure strategy is max-random. The linear strategy is next and even is the most insecure. The reason that the max-greedy/max-random strategy is more secure than linear/even has been explained in detail in \cite{KerBatchRealWorld}. In brief, this effect is caused by whitening the features in the pre-processing stage. Overall, with relatively large payloads, the LOF method achieves efficient detection results.

\begin{figure}
  \centering
  \includegraphics[width=4.5in]{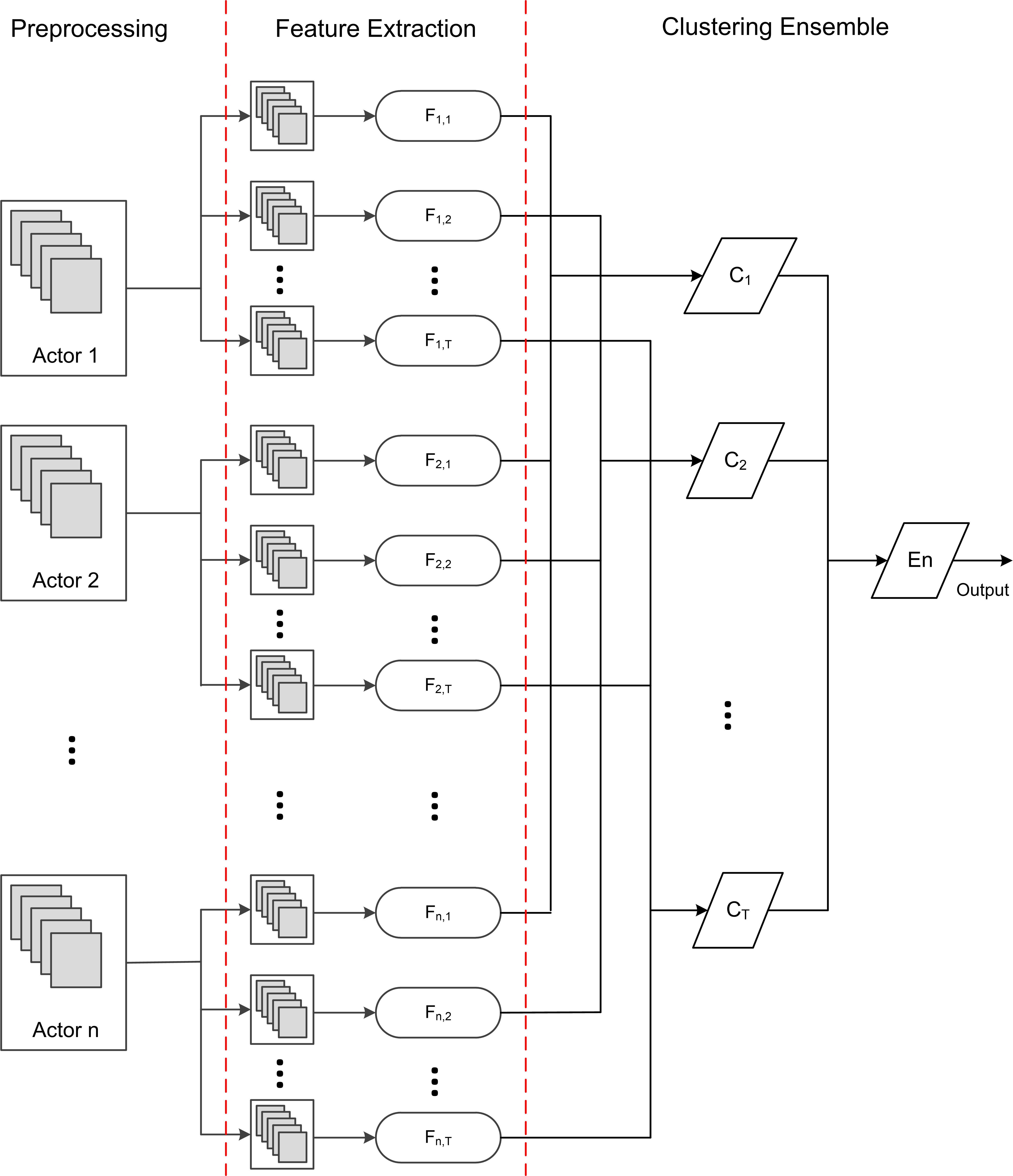}\\
  \caption{General framework for clustering ensemble.}\label{fig1-6}
\end{figure}

\section{Ensemble and Dimensionality Reduction}
\subsection{Clustering Ensemble}
The clustering ensemble method combines the detection results from multiple clustering sub-models. As shown in Figure 1.6, the clustering ensemble framework is comprised of three parts: preprocessing, feature extraction, and clustering ensemble. The preprocessing procedure requires us to analyze the characteristics of the steganographic algorithm so that those sensitive components concealing the steganographic traces can be separated out from the stego. The preprocessing could be cropping, high-pass filtering, and other operations that can improve the statistical discriminability between the steganographic noise and the cover content. It can be said that, the goal of preprocessing is to enlarge the ``signal-to-noise'' ratio.

After preprocessing, we extract steganalysis features from the preprocessed objects, which reduces the preprocessed objects to vectors with a small dimension. We use $T$ clustering sub-models $\{C_1, C_2, ..., C_T\}$ for clustering ensemble. For each sub-model $C_i$, we can directly apply the hierarchical clustering method, by which we can collect suspicious actor(s) that will be judged as the steganographer candidate(s). By merging the results from these sub-models, the index value(s) indicating the most suspicious actor(s) that will be chosen as the steganographer(s), can be determined.

For better understanding, we review Li \emph{et al.}'s work \cite{Li2016}, which is a related work exploiting clustering ensemble for SIP. In their work, the images held by actors are randomly cropped with smaller size to build actor subsets. Then, the LI-250 features mentioned in the aforementioned section are extracted from these actor subsets. After performing the agglomerative clustering in the (normalized) feature space, a most suspicious actor can be separated from the innocent ones. By repeating these steps, multiple decisions are made, and the final guilty actor can be identified by majority voting. It is mentioned that, in Li \emph{et al.}'s work, the clustering operation is slightly different from the above basic agglomerative clustering algorithm, and they use Euclidean distance for two feature vectors. We detail the algorithm below.

Denote a set of $n$ actors as $\{A_1, A_2, ..., A_n\}$, which include one guilty actor and $n-1$ normal (innocent) actors, each of whom transmits $m$ JPEG images. Clustering ensemble is used to identify the steganographer according to the following steps.
\begin{enumerate}
  \item Denote the image sets of $n$ actors as $\{I_1, I_2, ..., I_n\}$. Each $I_i~(1\leq i\leq n)$ contains $m$ JPEG images of size $M\times N$. Randomly crop each JPEG image in $I_i~(1\leq i\leq n)$ to size $m'\times n'~(m' < M, n' < N)$. The cropped images consist of $n$ subsets $I_1', I_2', ..., I_n'$, where $I_i'~(1\leq i\leq n)$ includes $m$ JPEG images of size $m'\times n'$.
  \item For each image subset $I_i'~(1\leq i\leq n)$, extract the LI-250 features to form feature sets $F_i~(1\leq i\leq n)$, where $F_i = (f_i^{(1)}, f_i^{(2)}, ..., f_i^{(m)})$ and each $f_i^{(j)}$ represents a 250-D feature vector. Notice that, each $F_i$ is associated with a second subscript $j~(1\leq j\leq T)$, i.e., $F_{i,j}$, to denote $F_i$ used in the $j$-th clustering sub-model since we have $T$ sub-models.
  \item Normalize the feature sets $\{F_1, F_2, ..., F_n\}$ such that every column of feature matrix has zero mean and unit variance. Then, the distance between different pairs of feature sets can be calculated with
            \begin{equation}
            D(F_i,F_j) = \frac{1}{m^2}\sum_{p=1}^{m}\sum_{q=1}^{m}d(f_i^{(p)}, f_j^{(q)}),i,j\in [1,n],
            \end{equation}
        where $d(f_i^{(p)}, f_j^{(q)})$ represents the Euclidean distance between the two feature vectors.
    \item Perform hierarchical clustering with the above distance measure. First, two actors with the minimum distance are combined into a new cluster, denoted by $\mathcal{X}$, and the remaining actors are in a set denoted by $\mathcal{Y}$. We repeat to select an actor from $\mathcal{Y}$ and add the selected actor into $\mathcal{X}$ until only one actor is left in $\mathcal{Y}$. At each selection step, the actor $Y\in \mathcal{Y}$ who has the smallest distance to $\mathcal{X}$ is selected, where the distance between an actor $Y\in \mathcal{Y}$ and a cluster $\mathcal{X}$ is defined as:
            \begin{equation}
            D(Y,\mathcal{X}) = \frac{1}{|\mathcal{X}|}\sum_{X\in\mathcal{X}}D(Y,X).
            \end{equation}
        The last remaining actor in $\mathcal{Y}$ is considered a suspicious actor.
    \item Repeat Steps 1-4 for $T$ times. Each time, we identify a suspicious actor, denoted by $C_i,~i=1,2,...,T$. By majority voting with $T$ sub-models, the actor who is identified as suspicious the most frequently is determined as the final culprit. When there is a tie, randomly select one to break the tie.
\end{enumerate}

The detection performance is evaluated by the overall identification accuracy rate, calculated as the ratio between the number of correctly detected steganographic actors and the selected total number of steganographic actors. In Li \emph{et al}'s method, the original images are cropped to images with a smaller size. Li \emph{et al.} point that, there is generally an optimal range for $m'~(=n')$, namely, when the value of $m'$ is far away from the optimal range, the detection performance will become worse. This interesting phenomenon is explained as:  When the size of the cropped images is too large, there is not much diversity in the randomly cropped images, and as such, each sub-clustering tends to generate same result. The benefit of clustering ensemble may disappear. When the size of the cropped images is too small, the statistical features of the images become unstable, leading to poor detection in each sub-clustering.

Experiments reported in \cite{Li2016} show that LI-250 outperforms pervious popular JPEG steganalysis features PEV-274 \cite{PEV274}, LIU-144 \cite{liu144}, DCTR-8000 \cite{dctr8000}, and PHARM-12600 \cite{PHARM12600}, which is explained as: on the one hand, comparing with PEV-274 and LIU-144, LI-250 could well reveal the changes of cover elements caused by the tested steganographic algorithms; on the other hand, although DCTR-8000 and PHARM-12600 are conclusively more sensitive for supervised binary classification, they have a high dimension and contain a number of weak features, which may lead to worse performance for SIP \cite{Ker2012, KerChallenges}.

Usually, for single hierarchical clustering, better performance can be obtained when the images are with large sizes \cite{SquareRoot}. However, good performance with large image size in the single hierarchical clustering method does not imply good performance of clustering ensemble using the same large image size for Li \emph{et al.}'s method. That is, when the size of cropped images is too large, the diversity in the randomly cropped images is reduced, and thus multiple clustering rounds are likely to generate the same result. The benefit of clustering ensemble disappears. Therefore, it can be said that, the preprocessing operation may have significant impact on the detection performance.

Intuitively, the higher $T$, the higher the overall accuracy, but the longer the running time. In \cite{Li2016}, the overall accuracy is tending towards stability with an increasing $T$, and does not change seriously when $T$ is larger than a threshold. It is believed that, the cropped images have a lot of overlapped area when $T$ increases, while the diversity will be reduced implying that multiple sub-models may generate the same results.

It can be easily inferred that the clustering ensemble framework shown in Figure 1.6 can be applied to other detection methods (e.g., LOF-based), by making adjustments accordingly. We do not provide more discussion here.

\subsection{Dimensionality Reduction}
Different from clustering ensemble (which has good generalization in terms of technical framework), dimensionality reduction focuses on the process of reducing the number of random variables under consideration by obtaining a set of principal variables. There are two important approaches for dimensionality reduction: feature selection and feature projection.

\subsubsection{Feature Selection}
Feature selection attempts to select a feature subspace from the original full feature space. Feature selection is often used in domains where there are many features and comparatively few sample points. Feature selection simplifies the model, and thus results in a shorter running time. It can also avoid disasters due to the dimensional problem. For SIP, in mainstream works, the distances in high dimensional space between features of actors are determined to find the abnormal cluster or outlier corresponding to the steganographer. However, in the high dimensional space, the distances between feature points may become similar to each other, resulting in that, it is not easy to separate the abnormal points from the normal ones. To this end, Wu \cite{WuFB} uses a simple feature selection based ensemble algorithm for SIP. The method merges results from multiple detection sub-models, each of which feature space is randomly sampled from the raw full dimensional space. Unlike the basic framework, the images held by a single actor are divided into multiple disjoint sets, resulting in that, each actor is represented by multiple sets of feature vectors.

In detail, let $A=\{a_1,a_2,...,a_n\}$ and $S(a_i)=\{\textbf{I}_1^{(i)},\textbf{I}_2^{(i)},...,\textbf{I}_m^{(i)}\}$ respectively represent actors and the images held by actor $a_i$. A detector computes the \emph{preprocessed} features for each $a_i$, i.e., $F(a_i)=\{\textbf{f}_1^{(i)},\textbf{f}_2^{(i)},...,\textbf{f}_m^{(i)}\}$. All $F(a_i)~(1\leq i\leq n)$ are divided to disjoint sets with an identical size, i.e.,
\begin{equation}
P(a_i)=\bigcup_{j=1}^{p}{P_j(a_i)},(1\leq i\leq n),
\end{equation}
where $m=p\cdot q$ and $P_j(a_i)=\{\textbf{f}_{jq-q+1}^{(i)},\textbf{f}_{jq-q+2}^{(i)},...,\textbf{f}_{jq}^{(i)}\}$.

Accordingly, each $a_i$ can be represented by $p$ set of feature vectors. We call the $p$ sets as ``$p$ points''. We can collect a total of $p\cdot n$ points, each of which belongs to one of the $n$ actors. It is naturally assumed that, distances between an abnormal point and a normal point should be larger than that between two normal points. In other words, normal points are densely distributed while abnormal ones are sparsely distributed. For any two points, we can use MMD or other metrics to measure their distance. However, when a point has only one feature vector, we cannot use MMD since its value always equals zero. In this case, one may adopt the Euclidean distance or other suitable metrics. Therefore, by anomaly detection (e.g., LOF detection), a ranking list for the $pn$ points is determined according to their anomaly scores.

We use $pn$ triples $\{(u_i,v_i,w_i)\}_{i=1}^{pn}$ to denote the sorted information, where $u_1\geq u_2\geq ...\geq u_{pn}$ represent the anomaly scores. $v_i$ denotes the corresponding actor and $w_i$ is the point index, namely, we have $P_{w_i}(v_i)\in P(v_i)$. For each actor $a_i$, we can determine a fusion score below:
\begin{equation}
s(a_i) = \sum_{j=1}^{pn}\frac{(pn+1-j)\cdot \delta(v_j,a_i)}{p},(1\leq i\leq n),
\end{equation}
where $\delta(x,y)=1$ if $x=y$, otherwise $\delta(x,y)=0$. By sorting the fusion scores, we can generate the final ranking list, where the actor with the largest score will be the most suspicious.

In this way, we can construct a single anomaly detection system operated on the full feature space. Obviously, we can build $T$ sub-models $\{\mathcal{M}_1,\mathcal{M}_2,$ ..., $\mathcal{M}_T\}$, whose feature dimensions are $\{d_1,d_2,...,d_T\}$. Each $d_i~(1\leq i\leq T),$ is chosen from $[H/2,H-1]$, where $H$ is the dimension of the raw full feature space. It is seen that, each sub-model is exactly the same as the single anomaly detection system, except that the used features are randomly sampled from the full feature space. For each $\mathcal{M}_i$, we can collect a ranking list, denoted by $\textbf{r}_i$ = $\{r_{i,1}$, ..., $r_{i,n}\}$, and $r_{i,j}$ means the actor with the $j$-th largest anomaly score, i.e., $r_{i,1}$ is the most suspicious and $r_{i,n}$ is the least suspicious. Accordingly, by further processing $\{\textbf{r}_1,\textbf{r}_2,...,\textbf{r}_T\}$, the final fusion score for $a_i$ can be:
\begin{equation}
s_\text{F}(a_i) = \sum_{j=1}^{T}\frac{n+1-\sum_{k=1}^{n}[k\cdot \delta(r_{j,k},a_i)]}{T}.
\end{equation}
By sorting $\{s_\text{F}(a_1), ..., s_\text{F}(a_n)\}$, the actor with the largest score will be the most suspicious, and the smallest score corresponds to the least suspicious.

It is observed that the method has the ability to improve the performance, which, however, is not significant and could be affected by parameter settings. For performance optimization, one may design advanced feature selection algorithms for choosing efficient feature components for detection.

\subsubsection{Feature Projection}
Feature projection should be distinguished from feature selection. While feature selection returns a subset of the original features by somehow way, feature projection creates new features from functions of the original features. It is excepted that, the projections of the original high dimensional features exhibit maximal information about the class label. Same as the feature selection, feature projection also reduces the complexity to help avoid over-fitting. For feature projection, the feature transformation can be either linear or non-linear. E.g., the linear principle components analysis (PCA) \cite{PCA} performs a linear mapping of the raw features to a lower-dimensional space such that the variance of the features in the lower-dimensional representation is maximized. Kernel PCA \cite{kernelPCA}, as a non-linear extension of PCA, uses a kernel to enable the originally linear operations of PCA to be performed in a RKHS.

Pevn$\acute{\text{y}}$ and Ker \cite{KerChallenges} have provided profound study on feature projection for SIP. They limit themselves to linear feature projections because it is believed that the steganalysis problem is essentially linear ``in some sense'', e.g., experiments in \cite{blindSteganalysis} demonstrate that accurate estimators of payload size can be created as linear functions of feature vectors. As pointed by \cite{KerChallenges}, we cannot expect that all stego objects move in the same direction, certainly not when created with direrent stego algorithms, nor that cover objects begin from nearby points if they arise from different sources. The boundary between cover and stego objects may be nonlinear. Accordingly, a good set of condensed features will be a collection of different linear projections, which between them capture the behaviour of different cover objects, sources, and embedding algorithms.

Therefore, to achieve better performance, they exploit principal component transformation (PCT), maximum covariance transformation, ordinary least square regression, and calibrated least squares regression for feature projection. The PCT can be defined as an iterative algorithm, where in $k$-th iteration one seeks a projection vector $w_k\in \mathbb{R}^d$ best explaining the raw features and being orthogonal to all projections $\{w_i\}_{i=1}^{k-1}$. It is formulated as:
\begin{equation}
w_k = \underset{||w||=1}{\text{arg~max}}~~~w^\text{T}\textbf{X}^\text{T}\textbf{X}w,
\end{equation}
subject to $w^\text{T}w_i=0,\forall i\in [1,k-1]$. $w_k$ is $k$-th eigenvector of $\textbf{X}^\text{T}\textbf{X}$, assuming that the vectors are sorted such that the eigenvalue sequence is non-increasing.

A weakness of PCT is that, it does not take into account the objective function. Pevn$\acute{\text{y}}$ and Ker define the object function as the presence of steganography signalled by the steganographic change rate. It is to some extent resolved by finding a direction maximizing the covariance between the projected data $\textbf{X}^sw$ and the dependent variable $\textbf{Y}^s$. Also, $w_k$ should be orthogonal to $\{w_i\}_{i=1}^{k-1}$. It is called the Maximum CoVariance (MCV) transformation method, which can be formulated as:
\begin{equation}
w_k = \underset{||w||=1}{\text{arg~max}}~~~{\textbf{Y}^s}^\text{T}\textbf{X}^sw,
\end{equation}
subject to $w^\text{T}w_i=0,\forall i\in [1,k-1]$. The analytical solution is $w_k = {\textbf{Y}^s}^\text{T}\textbf{X}_k^s$, where $\textbf{X}_k^s = \textbf{X}_{k-1}^s(\textbf{I}-w_{k-1}w_{k-1}^\text{T})$, and $\textbf{X}_1^s = \textbf{X}^s$.

Ordinary Least Square regression (OLS) finds a single direction $w$ minimizing the total square error between $\textbf{X}^sw$ and $\textbf{Y}^s$. Such optimization problem can be formulated as:
\begin{equation}
w = \underset{w\in \mathbb{R}^d}{\text{arg~min}}~~~2{\textbf{Y}^s}^\text{T}\textbf{X}^sw - w^\text{T}{\textbf{X}^s}^\text{T}\textbf{X}^sw,
\end{equation}
the analytical solution of which is:
\begin{equation}
w = ({\textbf{X}^s}^\text{T}\textbf{X}^s)^{-1}{\textbf{X}^s}^\text{T}\textbf{Y}^s,
\end{equation}
which assumes that, ${\textbf{X}^s}^\text{T}\textbf{X}^s$ is regular and the inversion is numerically stable. In practice, this is not always true, and steganalysis features may lead to a nearly singular matrix. To alleviate this, a small diagonal matrix is added to prevent non-singularity and increase the stability of the solution:
\begin{equation}
w = ({\textbf{X}^s}^\text{T}\textbf{X}^s + \lambda\textbf{I})^{-1}{\textbf{X}^s}^\text{T}\textbf{Y}^s.
\end{equation}

Good steganographic projections should be sensitive to embedding changes yet insensitive to the object content, meaning that, the covariance between the projection of stego features and their embedding change rate should be high, while the variance of projection of cover features should be low. This objective is not optimized in any of the above methods. To this end, Pevn$\acute{\text{y}}$ and Ker propose a so-called \emph{calibrated least squares} (CLS) regression method, which finds the projections by iteratively solving:
\begin{equation}
w = \underset{w\in \mathbb{R}^d}{\text{arg~min}}~~~2{\textbf{Y}^s}^\text{T}\textbf{X}^sw - w^\text{T}{\textbf{X}^c}^\text{T}\textbf{X}^cw-\lambda ||w||^2,
\end{equation}
subject to $w^\text{T}w_i=0,\forall i\in [1,k-1]$. And, the analytical solution is:
\begin{equation}
w_k = ({\textbf{X}_k^c}^\text{T}\textbf{X}_k^c+\lambda\textbf{I})^{-1}{\textbf{X}_k^s}^\text{T}\textbf{Y}^s,
\end{equation}
where $\textbf{X}_k^s = \textbf{X}_{k-1}^s(\textbf{I}-w_{k-1}w_{k-1}^\text{T})$, $\textbf{X}_1^s = \textbf{X}^s$, and $\textbf{X}_k^c = \textbf{X}_{k-1}^c(\textbf{I}-w_{k-1}w_{k-1}^\text{T})$, $\textbf{X}_1^c = \textbf{X}^c$.

Instead of PEV-274, Pevn$\acute{\text{y}}$ and Ker use CF-7850 \cite{ensembleTIFS} and outlier detection \cite{Ker2012} for experiments. Experimental results show that, in terms of identification accuracy (i.e., the average rank of the only one guilty actor), the unsupervised PCT method is worst. The MCV method is inferior to OLS in most situations, and CLS is significantly superior to the others. The explanation can be described as follows: PCT focuses on the explanation of the variance in the data, which is not exactly on par with our goal, as the variance in features can be dominated by cover image content. MCV finds directions maximally correlated with the explained variable (change rate). Although this is better than PCT, the image content acting as noise is not suppressed. OLS maximizes the covariance of a projection with the payload and minimizes the variance of the projection. Since the covariance and variance of projections are measured on the same data, it is not clear whether the variance comes from image content or from embedding changes, whereas, CLS removes this ambiguity.

\section{Conclusion and Further Research}
In this chapter, we have reviewed the problem of steganographer identification, in which we aim to identify a steganographer who sends many images (some of them may be innocent) in social networks or public environment of many other innocent users. The analyzer must deal with multiple users and multiple images per user, and particularly the differences between cover sources, as well as the differences between the covers and the stegos.

In despite of being posed for more than ten years, this problem was first addressed by Ker and Pevn$\acute{\text{y}}$ \cite{Ker2011} in 2011. They introduced two general technical frameworks for steganographer identification, i.e., clustering-based detection and outlier-based detection. It can be said that, most of the advanced works reported in the literature are based on them. Though, throughout this chapter, we consider only one guilty actor (i.e., the steganographer), we can actually identify a short list of $k$ most suspicious actors.

It has been shown that, both ensemble and dimensionality reduction can improve the identification performance. As a common operation used in steganalysis, ensemble merges results from multiple sub-models to make a final decision. Dimensionality reduction, seeming to be more important, exploits salient features and eliminates irrelevant feature fluctuations by representing the discriminative information in a lower dimensional manifold \cite{infoFSP}.

The outlier-based analyzer is not compatible with rich features containing lots of weak features, because the weak features contain lots of noise (caused by cover content), resulting in inferior performance. Pevn$\acute{\text{y}}$ and Ker \cite{KerChallenges} used supervised feature reduction methods to improve performance, meaning that, the complete outlier detector was no longer unsupervised. However, the features showed a high level of robustness to varying the embedding algorithm, meaning that the outlier detector retained its universal behaviour. Finding the balance between supervision (to focus information) and universality (for robustness) is an important challenge for pooled steganalysis for SIP, which needs further investment.

Building upon advances in machine learning, Pevn$\acute{\text{y}}$ and Nikolaev \cite{optFunction} attempted to learn an optimal pooling function for pooled steganalysis. Although experiments show that learned combining functions are superior to the prior art, many interesting phenomenons were found, pointing to directions of further research, e.g., are there better strategies for the steganographer to distribute the message into multiple objects, or the linear strategy is the best he can do despite the steganalyzer knowing it? It is believed that, more questions will arise as the study of SIP moves ahead \cite{move2real}.

\end{document}